\documentclass[aps,preprint, onecolumn, superscriptaddress]{revtex4}
\usepackage{graphicx,epsfig}
\usepackage{epstopdf}
\DeclareGraphicsExtensions{. jpg,. pdf, . mps, .png, .eps, . ps, . EPS}

\AppendGraphicsExtensions{.tif}
\usepackage{amsmath}
\usepackage{color}

\usepackage{amsthm}
\usepackage{amssymb}
\def\be{\begin{equation}}
\def\ee{\end{equation}}

\def\bc{\begin{center}}
\def\ec{\end{center}}
\def\bea{\begin{eqnarray}}
\def\eea{\end{eqnarray}}

\begin{document}
\title{Emergent Hyperbolic Network Geometry}

\author{Ginestra Bianconi} 

\affiliation{School of Mathematical Sciences, Queen Mary University of London, E1 4NS London, United Kingdom.}
 \author{Christoph Rahmede }
  \affiliation{Rome International Centre for  Material Science Superstripes RICMASS, via dei Sabelli 119A, 
00185 Roma, Italy}

\begin{abstract}
{\bf
A large variety of interacting complex systems are characterized by  interactions occurring between more than two nodes. These systems are  described by simplicial complexes. Simplicial complexes are formed by simplices (nodes, links, triangles, tetrahedra etc.) that have a natural geometric interpretation. As such simplicial complexes are widely used in quantum gravity approaches that involve a discretization of spacetime.  
Here, by extending our knowledge of growing complex networks to growing simplicial complexes we investigate the nature of the emergent  geometry of complex networks and explore whether  this geometry is  hyperbolic.  Specifically we show that an hyperbolic network geometry emerges spontaneously from models of growing simplicial complexes that are purely combinatorial. The statistical  and geometrical properties of the growing simplicial complexes strongly depend on their dimensionality and display  the major universal properties of real complex networks (scale-free degree distribution, small-world and communities) at the same time.  Interestingly, when the network dynamics includes an heterogeneous fitness of the faces, the growing simplicial complex can undergo phase transitions that are reflected by  relevant  changes in the network geometry. 
}\end{abstract}
\maketitle

\section*{INTRODUCTION}
Simplicial complexes are the  many-body generalization of networks  \cite{BA,SW,Doro_book,Newman_book,Santo,Laszlo_book} and they can encode interactions occurring between two or more nodes  \cite{Bassett,flavor,Kahle,CQNM,Emergent,Newman_hypergraph,Dima,Owen}. While  networks are formed exclusively  by nodes and links, simplicial complexes include  higher dimensional simplices i.e.  triangles, tetrahedra etc. As such they  are fundamental  to study a large variety of real complex interacting systems, including brain functional networks \cite{Bassett}, protein interaction networks   \cite{proteins}, collaboration networks    \cite{Newman_hypergraph}.    Because simplices have a natural topological and geometrical interpretation, simplicial complexes are ideal  to investigate the underlying  geometry and topology of networks   \cite{perspective,Bassett,Vaccarino1,Vaccarino2} and for these reasons they are extensively used in quantum gravity  \cite{Lee,CDT1,CDT2,Oriti,Cortes}. 

One of the fundamental quests of quantum gravity is to describe the emergence of a continuous, finite dimensional space, using pre-geometric models, where space is an emergent property of a network  or of a  simplicial complex  \cite{pregeometry_review,CDT1,CDT2,graphity}. This fundamental mathematical problem has its relevance also in the field of  network theory \cite{perspective} where one of  the major aim of network geometry is to characterize the continuous hidden metric behind the inherently discrete structure of complex networks. In fact, it is  believed that most complex networks have a continuous hidden network geometry \cite{Boguna_Internet,Boguna_navigability, Boguna_metabolic,Boguna_growing,Boguna_hyperbolic} such that any two  connected  nodes are also close in the hidden metric   \cite{Kleinberg,Aste,Boguna_navigability,Boguna_Internet}. In this context, there  is increasing evidence that the hidden geometry of a large variety of networks including  the Internet,  airport networks, the brain  functional networks, and metabolic networks   \cite{Boguna_Internet,Boguna_navigability, Boguna_metabolic,Boguna_growing} is hyperbolic.   Characterizing the hyperbolicity of  networks is not only a fundamental theoretical question,  but it can also have practical implications as it can be used to improve significantly the navigability on such networks   \cite{Boguna_navigability,Boguna_Internet}. While the mathematical definition of  the curvature of networks is a hot mathematical subject for which different  definitions have been given   \cite{perspective,Yau1, Yau2, Gromov,Jost1,Jost2}, most of the results obtained so far are  related to  the  embeddings of complex networks in hyperbolic spaces   \cite{Aste,Boguna_navigability,Boguna_Internet,Boguna_metabolic,Boguna_growing}.

The underlying assumptions of  several models  \cite{Boguna_hyperbolic,Boguna_growing}  of complex hyperbolic networks is that nodes   are sprinkled randomly in the hidden hyperbolic metric and links are established according to their hyperbolic distance. Interestingly this type of  models can be related  to causal sets \cite{Sorkin} in  de Sitter space and they have been used to describe  a "network cosmology" \cite{Network_cosmology}.  From the complexity point of view, if we want  for example to use this type of  models for describing the evolution  of the World-Wide-Web, the  sprinkling of the nodes in hyperbolic space reflects some distribution of interest of the webpage owners, and links between the webpages  are established  depending on the similarities between the  interests of webpage owners.
 Although this is a very plausible mechanism for network evolution, it  cannot be adopted if  we aim at  describing the  emergence of the underlying hyperbolic geometry  as the result of the endogenous dynamics of the network.

Here we  will show in the framework of  a very simple, stylized model,  that the hyperbolic network geometry can be  an emergent property of growing simplicial complexes that  share  the universal properties of  complex network structures. Specifically,  we will propose a model in which  the hidden hyperbolic  metric is not causing the network dynamics but it is instead  the outcome of the network evolution.

 Our model of emergent geometry is based on a growing simplicial complex.  Metric spaces  satisfy the triangular inequality, therefore a network with non-trivial geometry  should include a high clustering coefficient and high density of triangles, ensured by building the network using simplicial complexes. Additionally growing networks have been extensively used as  a non-equilibrium framework   \cite{BA,Doro_book,Doro_model,Krapivsky,BB,BE} for  the emergence of complex statistical  properties of  networks such as the power-law degree distribution. 
By extending the well established framework of growing network models \cite{BA,Doro_book,Doro_model,Krapivsky,BB,BE}   to simplicial complexes we will provide significant new insights into emergent geometry.
Importantly,  the emergent hidden geometry of growing simplicial complexes  is  hyperbolic, i.e. the hyperbolic geometry  emerges spontaneously from the evolution of the simplicial complexes. In this way we provide evidence that hyperbolic network geometry emerges from growing simplicial complexes whose temporal evolution is purely combinatorial, i.e. it does not take into account the hidden geometry.

Interestingly the properties of the network geometry  change significantly with the dimension of the simplicial complex and the network geometry can be  strongly affected by phase transitions occurring when a fitness parameter \cite{BB,BE,CQNM,flavor} is associated to each  face of the simplicial complexes describing intrinsic local heterogeneities.

\section*{RESULTS}
We consider simplicial complexes formed by gluing together $d$-dimensional simplices. A $d$-dimensional simplex (or $d$-simplex)  is a topological object including the set of $d+1$ nodes and all its subsets.
The  underlying network structure of a $d$-simplex is constituted  by a fully  connected network,  or a clique,  of $d+1$ nodes,  such as links ($1$-simplices), triangles ($2$-simplices),  tetrahedra ($3$-simplices) etc. The  $\delta$-faces of a $d$-dimensional simplex are all the $\delta$-dimensional simplices that can be built by a subset of $(\delta+1)$ of its nodes.  For example the faces of a triangle (2-simplex) are its three links (1-simplices) and its three nodes (0-simplices), the faces of a tetrahedron are its four triangular faces (2-simplices), its six links (1-simplices) and its four nodes (0-simplices), etc.
As long as we are concerned exclusively with the network properties of simplicial complexes, the use of simplicial complexes is equivalent with the use of hypergraphs and hypernetworks that are recently attracting increasing attention \cite{Newman_hypergraph,Coutinho}.

The simplicial complexes that we are considering in this paper are constructed by gluing $d$-simplices along their $(d-1)$-faces.
To every $(d-1)$-face $\alpha$ of the simplicial complex, (i.e. a link for $d=2$, or a triangular face for $d=3$) we associate an {\em incidence number} $n_{\alpha}$ given by the number of $d$-dimensional simplices incident to it minus one. 
The simplicial complex dynamics is dictated by the following algorithm and depends on a parameter $s=-1,0,1$ called {\em flavor}.  We start from  a single $d$-dimensional  simplex, i.e a triangle for $d=2$, a tetrahedron for $d=3$. At each time we add a  $d$-dimensional simplex to a $(d-1)$-face $\alpha$. The face $\alpha$ is chosen randomly with   probability  $\Pi_{\alpha}$ given by 
\bea
\Pi_{\alpha}=\frac{1+s n_{\alpha}}{\sum_{\alpha'}1+s n_{\alpha'}}.
\label{prob}
\eea
The new $d$-dimensional simplex is induced  by a new node and all the nodes of the chosen $(d-1)$-face $\alpha$.
For this type of dynamics, the  combinatorial condition to obtain a discrete manifold is that $n_{\alpha}$ can take exclusively the values $n_{\alpha}=0,1$. 

For $s=-1$ it is possible to attach a simplex only to faces with $n_{\alpha}=0$.
In fact for $n_{\alpha}=0$ we have  $\Pi_{\alpha}=\frac{1}{\sum_{\alpha'}1+s n_{\alpha'}}$ but for $n_{\alpha}=1$ we have $\Pi_{\alpha}=0$. As a consequence of this, the resulting network is a manifold, with each $(d-1)$-face incident at most  to two $d$-dimensional simplicial complexes, i.e. $n_{\alpha}=0,1$. 
For $s=0$ the $d-$dimensional simplices are attached with uniform probability to any $(d-1)$-face, while for $s=1$ the dynamics follows a generalized preferential attachment and the new simplex is attached to a $(d-1)$-face $\alpha$ proportionally to the number of simplicies already attached to the face , i.e. $1+n_{\alpha}$.
Therefore for  $s=0$ as for $s=1$ the incidence number $n_{\alpha}$ can  take values $n_{\alpha}=0,1,2,3\ldots$.

Simplices  are topological objects that can be turned into geometrical entities when we attribute a given length to their links.
Here, in order to describe the emergent geometry of our model of growing simplicial complexes, we assume  that every simplicial complex is built by  simplices that have  links of equal  length across the entire simplicial complex. 

The resulting networks are small world for every flavor $s$ and any dimension $d$ except from the special case $s=-1,d=1$ in which the resulting network is a chain. This implies that the number of nodes in the network $N$ increases exponentially with its diameter $D$, i.e. $N\simeq e^{D}$. 
Therefore, if all the links have equal  length,  the hidden geometry of these networks cannot be the one of a  Euclidean  space of finite Hausdorff dimension $d_E$ because this would imply a power-law scaling $N\simeq D^{d_E}$.
As a consequence of this the small-world property  suggests that  the natural embedding of these networks is  hyperbolic. 
Nevertheless the small-world property might not be sufficient to guarantee an embedding in the hyperbolic space. 
Here we show that for our class of growing simplicial complexes the  hidden geometry,  corresponding to the embedding where all the links have the same distance, are the hyperbolic spaces ${\mathbb {H}}^d$, and specifically the Poincar\'{e} ball model \cite{Hyperbolic}.   

The great advantage of the present class of models with respect to general small-world networks is that their dual is a tree. The dual network  can be constructed  by associating  to every $d$-simplex a node of the dual network, and to every  pair of $d$-simplicies sharing a $(d-1)$-face a link of the dual network. Since in our simplicial complex evolution at each time we glue a new $d$-simplex to a $(d-1)$-face,   the resulting structure of the dual network is a tree.
Taking advantage of this simple structure of the dual the present class of models admits several embeddings in the ${\mathbb{H}}^d$   hyperbolic space model. 
Between the different possible embeddings, only one embedding can  fill the entire space in the asymptotic limit $t\to \infty$. Therefore this embedding defines the emergent geometry of our simplicial complexes.
 
\begin{figure}
\includegraphics[width=15cm]{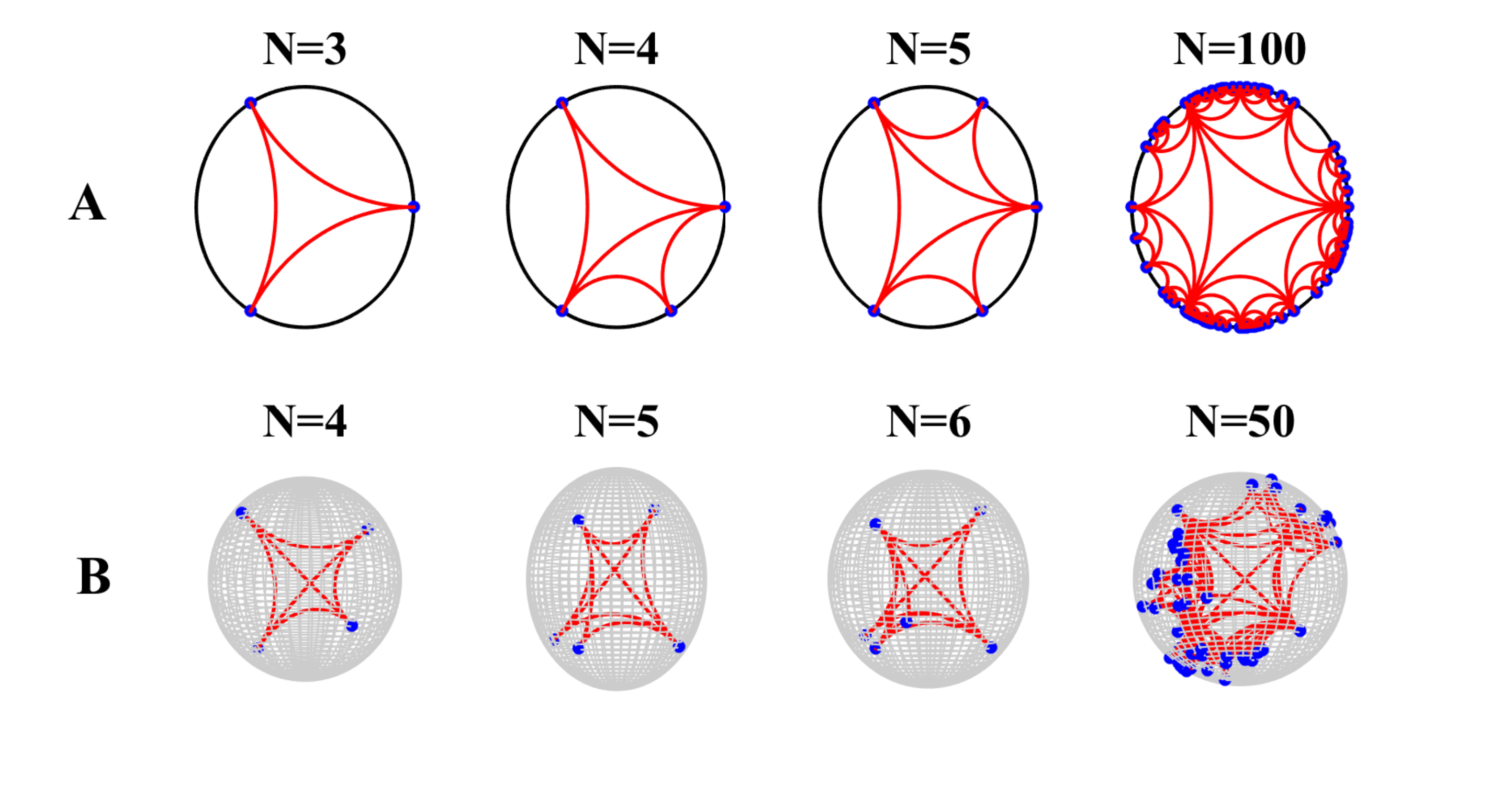}
\caption{ The first steps of the temporal evolution of growing simplicial complex with $N$ nodes is shown here in the emergent hyperbolic space for   $d=2$ (panel A) and $d=3$ (panel B). The flavor is  $s=-1$.}
\label{Figure1}
\end{figure}

Let us   consider a Poincar\'e ball model of $\mathbb{H}^d$. The Poincar\'e ball model includes all the points of the unit ball $B^n=\{{\bf x}\in\mathbb{R}^d: |{\bf x}|<1\}$, with $|\ldots|$ indicating the Euclidean norm. The Poicar\'e ball model is associated to the hyperbolic metric $d_B$ assigning to  each pair of points ${\bf x,y}\in \mathbb{R}^d$ the distance 
\bea
\cosh d_B({\bf x, y})=1+\frac{2|{\bf x-y}|^2}{(1-|{\bf x}|^2)(1-|{\bf y}|^2)}.
\eea
Here we identify every   $d$-dimensional simplex of our simplicial complex with an  ideal simplex of the Poincar\'e ball model. An ideal simplex  has all its  nodes  at the boundary of the hyperbolic ball, so all the nodes $i$ have a position ${\bf r}_i\in \mathbb{R}^d$ satisfying $|{\bf r}_i|=1$. This allows  to have all the nodes of each simplex at equal hyperbolic distance. Note that interestingly this distance is actually infinite but this is the "cost" required for having an embedding that asymptotically in time fills the entire hyperbolic space.   In order to fully characterize  the  hidden geometry of the studied networks, we need also to determine further  the  position of the ideal nodes at  the boundary of the  ball. To this end we start from a $d$-dimensional simplex whose ($d+1$) nodes have the same maximum (Euclidean) distance from each other. Therefore  the positions ${\bf r}_i$ of the initial $(d+1)$ nodes $i=1,2,\ldots d+1$ satisfy 
\bea 
\sum_{i=1}^{d+1}{\bf r}_i={\bf 0}.
\eea
Each new node $i$ of the network has a position $\bf{r}_i\in \mathbb{R}^d$ at the boundary of the ball determined  by the position of its "ancestors", i.e. the nodes  of the  face $\alpha$ connected to the new node. In particular the new node $i$ is placed at equal (Euclidean)  distance from all the nodes $j$ of the $\alpha$ face to which it is attached,i.e.
\bea
{\bf r}_i=\frac{\sum_{j\subset \alpha}{\bf r}_j}{\left|\sum_{j\subset \alpha}{\bf r}_j \right|}.
\eea 
In this way the angular position of the new node is fully determined by the stochastic dynamics of the network (see Figure \ref{Figure1} for details).
The resulting networks have a rich geometrical structure, which is linked to the mathematics of Farey sequences in $d=2$ \cite{Farey}. Additionally, the simplicial complexes in dimension $d=3$  are characterized by a boundary with notable geometrical features.The  induced geometry on this boundary can be studied by placing the nodes of the network in a $(d-1)$-dimensional space characterized by the angular coordinates of the nodes. The network resulting from the projection on the boundary  of the ball $B^n$ is a  random Apollonian network \cite{apollonian,apollonian2} for $d=3$.

Let us make three important observations related to the geometric nature of the proposed class of growing simplicial complexes.

First of all we note that the hyperbolic nature of the emergent geometry it is a  consequence of the assumption that each link of the simplicial complex must have equal length. This assumption implies that distances of different links can be compared. Therefore strictly speaking here the network geometry is actually a  consequence of a kind of "proto-geometry" that allows comparison of length of different links.
If we allow, instead, to have links of different lengths the curvature of the hidden geometry is not determined and it is even possible to tune the length of the links such that the same simplicial complex can be embedded in a $d$-sphere. For instance this can be achieved simply by   taking  the embedding on the Poicar\'e ball model described above, and considering  instead of the  hyperbolic metric on the ball the Euclidean metric.

Secondly we note  that the natural hyperbolic embedding of growing  simplicial complexes that we discussed above, works particularly well for  flavor $s=-1$ while some caution is required  when using this embedding for flavors $s=0$ and $s=1$. In fact, for $s=-1$ the simplicial complexes  are   manifolds and as a consequence of this,  links do not cross and each node of the simplicial complex has a distinct position in the hyperbolic space. However for flavor $s=0$ and $s=1$ the proposed embedding implies that some nodes of the simplicial complex  (the nodes that are immediate "descendant" of the same face) acquire the same position in the hyperbolic space. As a result in a geometrical embedding  links are effectively weighted. For this reason the growing simplicial complexes with flavor $s=-1$ play a very special role with respect to the other flavors $s=0$ and $s=1$.

Finally  we note  that for $s=-1$  and $d=3$ the growing simplicial complex model presented here belongs to the  class of stacked polytopes that are equivalent to Apollonian packings, whose  discrete  Lorentzian geometry  is raising recent interest in the mathematical community \cite{apollonian2,Apollonian_group,Chen_Hao1,Chen_Hao2}.
In fact, these stacked polytopes in $d=3$ have a symmetry group ${\cal G}$ that is a noncompact discrete subgroup of the Lorentz group  SO(3,1)=SL(2,$\mathbb{C}$)/$\mathbb{Z}_2$. Therefore these results provide an additional important insight on the  hyperbolic nature of the underlying geometry  of the class of  models proposed in this paper.

\begin{figure}
\includegraphics[width=16cm]{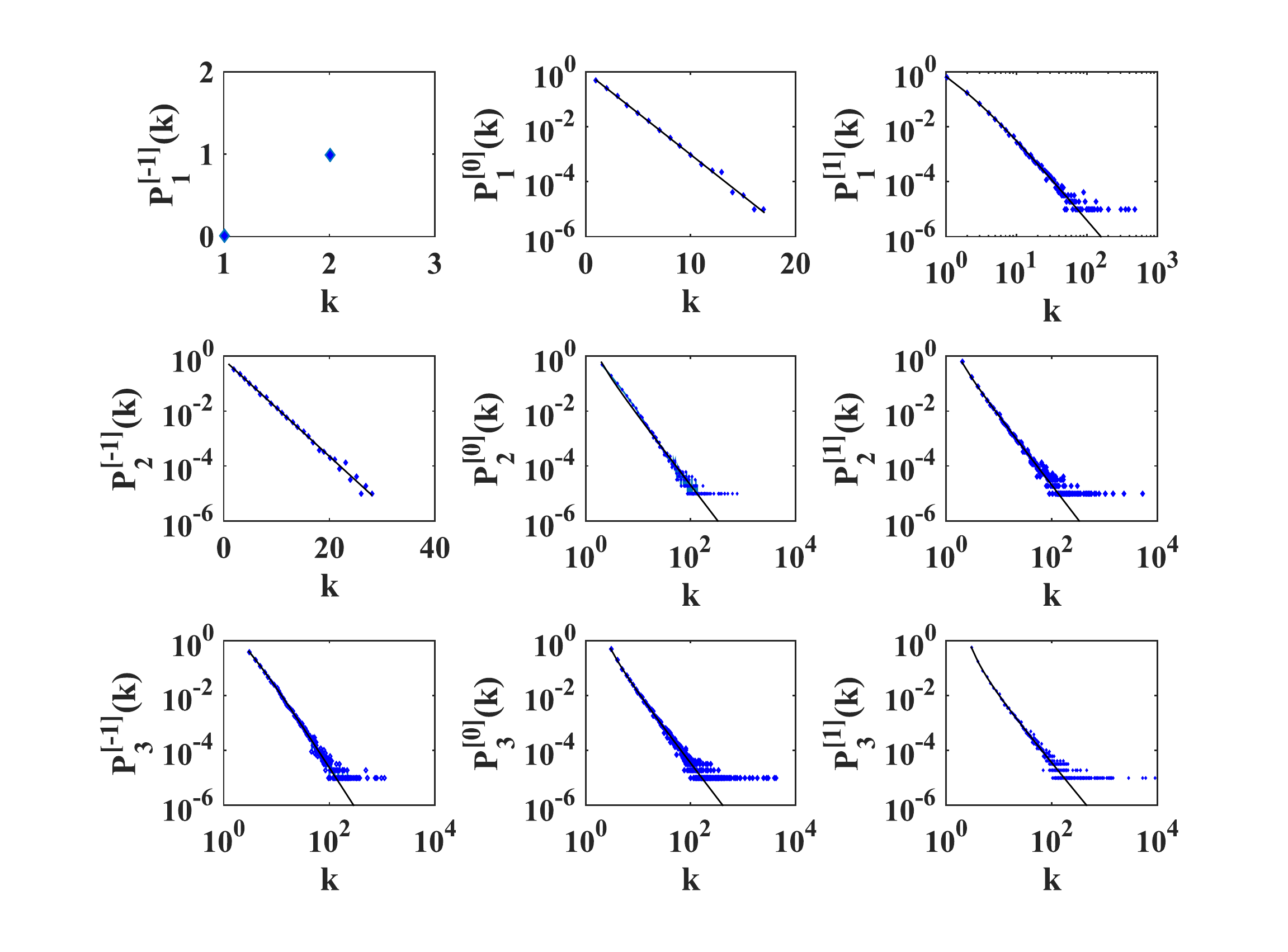}
\caption{{ The effect of dimensionality and flavor in the degree distribution $P_d^{[s]}(k)$. The symbols (blue dimonds) indicate the simulation result of a single realization of the growing simplicial complex with dimension $d$ and flavor $s$ with $N=10^5$ nodes. The solid lines indicate the theoretical predictions.} }
\label{Figurepk}
\end{figure}
\begin{table}
\parbox{.45\linewidth}{
\centering
\begin{tabular}{@{}cccc}
\hline
$M$ &$s=-1$&$s=0$&$s=1$\\
\hline
$d=2$&0.97&0.94&0.90\\
\hline
$d=3$&0.91&0.85& 0.80
\end{tabular}
}
\hfill
\parbox{.45\linewidth}{
\centering
\begin{tabular}{@{}cccc}
\hline
 $C$&$s=-1$&$s=0$&$s=1$\\
\hline
$d=2$&0.65&0.74&0.79\\
\hline
$d=3$&0.77&0.81& 0.84
\end{tabular}\\
}
\caption{ \label{table1}Modularity $M$ and average clustering coefficient $C$ of the growing simplicial complex with $N=10^4$ nodes  averaged over 20 realizations are reported here for the dimension $d=2,3$ and flavor $s=-1,0,1$. The modularity $M$ is  obtained by running the generalized Louvain method \cite{Gen_louvain,Louvain}.}
\end{table}
The  networks resulting from the proposed model of growing   simplicial complexes  are strongly affected by their dimensionality. In fact they are scale-free for dimension 
\bea
d>(1-s),
\eea
while for dimension $d=1-s$ they have exponential degree distribution.  
 In particular, the degree distribution $P_d^{[s]}(k)$ of growing simplicial complexes of dimension $d$ and flavor $s$ is given for  
for $d+s=1$ by (see Methods for details)
\bea
P_{d}^{[s]}(k)&=&\left(\frac{d}{d+1}\right)^{k-d}\frac{1}{d+1}, 
\eea
with  $k\geq d$ while for $d+s>1$ it is given by  (see Methods for details)
\bea
P_{d}^{[s]}(k)&=&\frac{d+s}{2d+s}\frac{\Gamma[1+(2d+s)/(d+s-1)]}{\Gamma[d/(d+s-1)]}\frac{\Gamma[k-d+d/(d+s-1)]}{\Gamma[k-d+1+(2d+s)/(d+s-1)]},
\label{Pksf0}
\eea
with $k\geq d.$\\
Therefore for $d+s>1$ the degree distribution is scale-free and has a power-law scaling 
\bea
P_{d}^{[s]}(k)\simeq k^{-\gamma_d^{[s]}}
\eea
for $k\gg 1$, with power-law exponent $\gamma_d^{[s]}$
\bea
\gamma_d^{[s]}=2+\frac{1}{d+s-1}\leq 3.
\eea
Finally for $d+s=0$ the simplicial complexes reduce to chains (see Methods for details).\\
In Figure $\ref{Figurepk}$ we show the perfect agreement between the predicted degree distribution and simulation results for $d=1,2,3$ and flavor $s=-1,0,1$.
  For dimensions $d>1$ these networks display a significant community structure (high modularity) and high average clustering coefficient (see Methods) as most complex networks.   The values of the modularity and the clustering coefficient are modulated by the dimension $d$ and the flavor $s$ of the growing simplicial complex (see Table $\ref{table1}$).

\begin{figure}
\includegraphics[width=12cm]{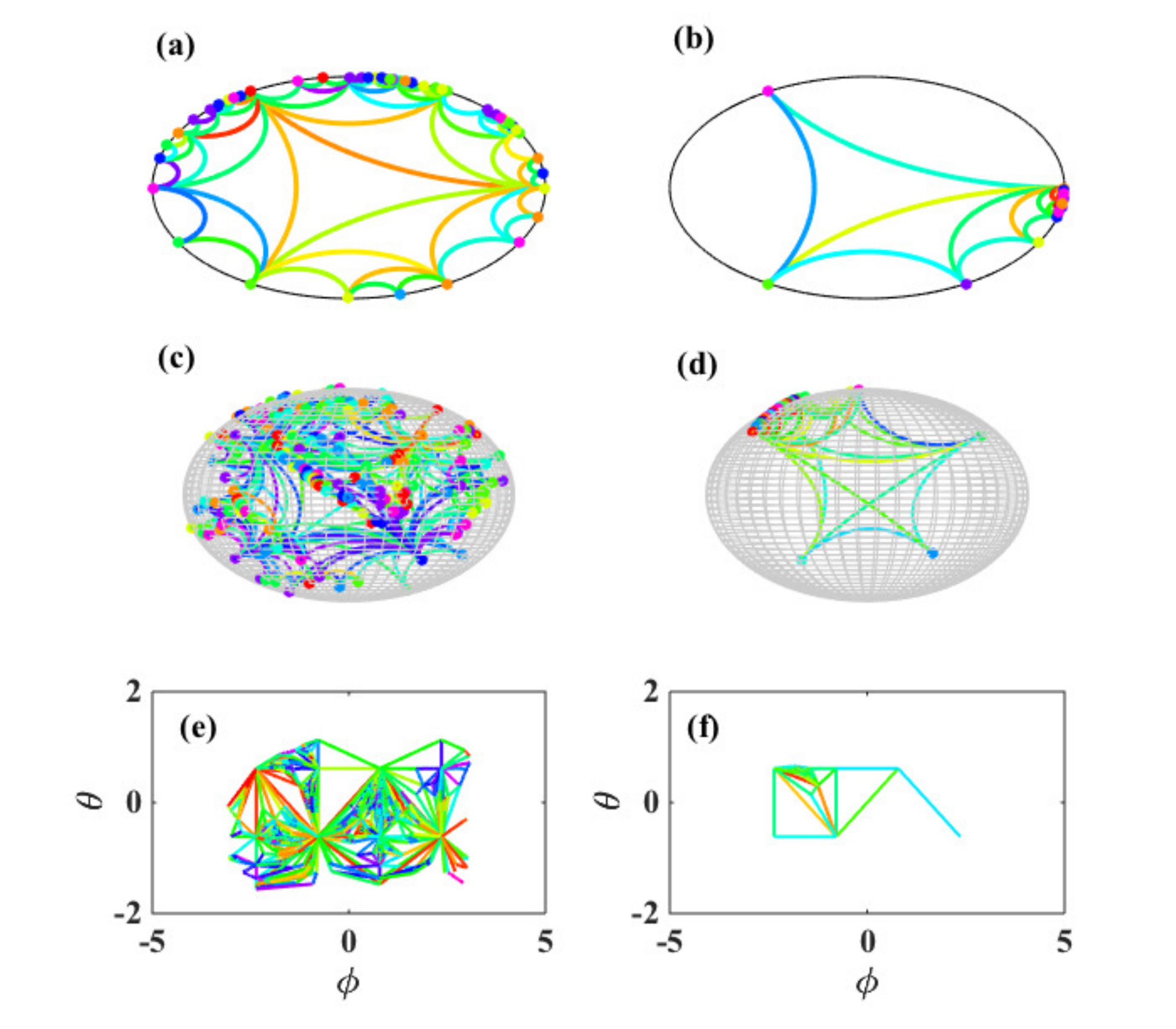}
\caption{The hyperbolic emergent network geometry is shown here for the growing simplicial complex with fitness of the nodes and flavor $s=-1$ above and below the phase transition in dimension $d=2$ (panel a-b) and dimension $d=3$ (panel c-d). Panel (e-f) display the projection of the network  on the boundary of the $d=3$ hyperbolic space. The energy distribution is uniform over discrete values of the energies of the nodes $0\leq\epsilon_i< 10$. The inverse temperature is $\beta=0.01$ for panels a,c,e, $\beta=50$ for panel b, $\beta=20$ for panels d,f. The number of nodes is $N=200$. The color of the links indicates the different values of their energies.}
\label{Figure2}
\end{figure}

The emergent geometry of growing simplicial complexes is strongly affected by phase transitions occurring in the network evolution. In order to show evidence for this statement we study a variation of the model including fitness of the faces of the simplicial complex.
First we associate to each node an energy $\epsilon_i$ drawn from a $g(\epsilon)$ distribution. To any $\delta$-face $\alpha$ with $0<\delta<d$ we associate an energy 
\bea
\epsilon_{\alpha}=\sum_{i\subset \alpha}\epsilon_i
\eea
and a fitness 
\bea
\eta_{\alpha}=e^{-\beta \epsilon_{\alpha}},
\eea
where $\beta=1/T>0$ is a global parameter called {\em inverse temperature}, such that for $\beta=0$ all the nodes have the same fitness while for $\beta \gg 1$ small differences in energy yield big differences in the fitness of the nodes.
The model remains the same with the exception that the probability $\Pi_{\alpha}$ that a $(d-1)$-face $\alpha$ is selected is no longer given by Eq. $(\ref{prob})$ but is given by 
\bea
\Pi_{\alpha}=\frac{\eta_{\alpha}(1+s n_{\alpha})}{\sum_{\alpha'}\eta_{\alpha'}(1+s n_{\alpha'})}.
\label{prob2}
\eea
This model displays a structural phase transition at high value of $\beta$. In order to characterize this phase transition we make use of the hyperbolic embedding.
While above the phase transition the simplicial complex is growing in all directions of the hyperbolic space and has statistical properties related to quantum statistics \cite{CQNM,flavor}, below the phase transition there is  symmetry breaking and the network evolves asymmetrically in the hyperbolic space.
In figure $\ref{Figure2}$ we show a visualization of the model above and below the phase transition for dimension $d=2,3$ showing that also the geometry of the boundary of simplicial complexes in $d=3$ is  strongly affected by the geometrical phase transition occurring in the model.

 \begin{figure}
 \begin{center}
\includegraphics[width=9cm]{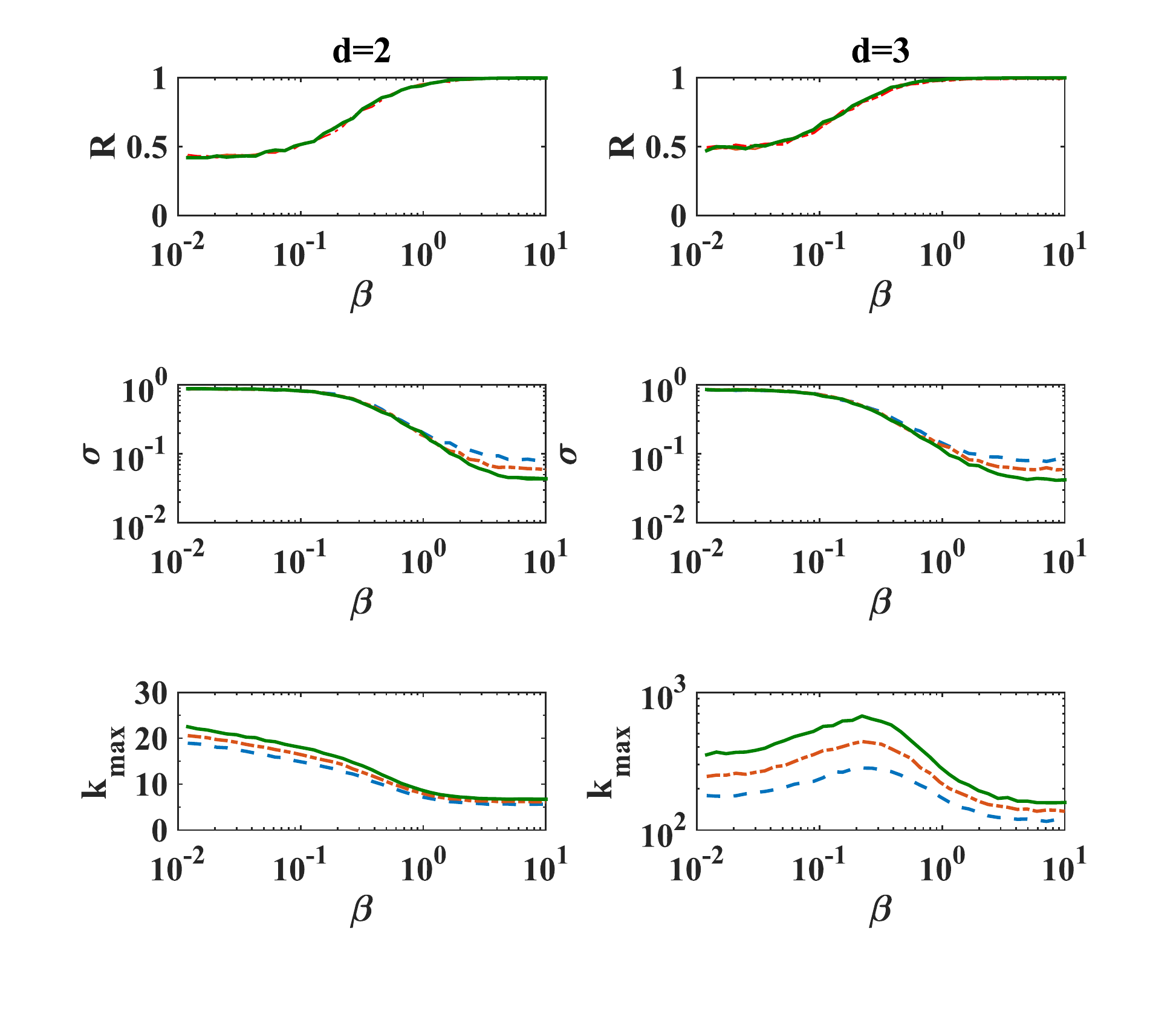}
\end{center}
\caption{ The phase transition in the network geometry of the growing simplicial complexes with fitnesses is characterized here by showing $R,\sigma$  and the maximum degree $k_{max}$ as a function of $\beta$. Here simulation results are reported for growing simplicial complexes with flavor   $s=-1$, dimension $d=2$ (left panels), and $d=3$ (right panels) and network sizes $N=2500$ (blue dashed line), $5000$ (red dot dashed line), $10000$ (green solid line). All the data are averaged over $500$ realizations.}
\label{Figure3}
\end{figure}
In order to numerically study the phase transition occurring in the growing simplicial complexes with fitnesses we define a vector ${\bf R}$ given by 
\bea
{\bf R}=\frac{1}{N}\sum_i {\bf r}_i,
\eea
where ${\bf r}_i$ is the (Euclidean) position vector of the node $i$ in the Poincar\'e ball.
We study  the Euclidean norm $R=|{\bf R}|$ and the standard deviation 
$\sigma=\sqrt{\frac{1}{N}\sum_i |{\bf r}_i-{\bf R}|^2},$
 and the maximum degree $k_{max}$ as a function of the inverse temperature $\beta$. Here we focus on  the result for $d=2,3$, flavor $s=-1$ (see Figure \ref{Figure3}). As $\beta$ increases across the phase transition, $\sigma$ develops relevant finite size effects and becomes vanishingly small in the large network limit. For the same values of the parameters $R$ approaches one indicating that the simplicial complex grows in a well defined preferential direction. The phase transition in the network geometry for flavor $s=-1$ has these characteristics both for $d=2$ and $d=3$. However the behavior of the  maximum degree $k_{max}$ across the phase transition shows major differences between  the case $d=2$ and $d=3$, displaying a clear maximum at the transition point for $d=3$.
 This implies that the transition might affect the degree distribution in different ways depending on the dimension $d$. Interestingly similar transitions are observed for different flavors $s=0,1$.
 
\section*{DISCUSSION}
 In conclusion,  this paper shows that the study of  simplicial complexes allows great advances in  our understanding of complex networks. In fact by extending the framework of growing complex networks to simplicial complexes our simple model produces networks that display most of  the universal properties of complex networks including scale-free degree distribution, small-world properties and significant modular structure.   These networks have statistical and geometrical properties that are a function of their dimension $d$ and their flavor $s$ that modulate the values of  their modularity and their clustering coefficient.
 These non-equilibrium models of simplicial complexes are ideal frameworks to show the  emergence of the hyperbolic network geometry.  Specifically  they can explain how real hyperbolic networks might result from purely combinatorial rules that do not take into consideration the hidden metric of the network.This network geometry has a very interesting structure linked to Farey sequences and Apollonian tilings. Additionally these network geometries can undergo relevant changes following phase transitions in the network evolution.

We believe that this paper, showing that growing simplicial complexes give rise to a complex emergent hyperbolic geometry, related to  Apollonian packings,  is opening new perspectives for understanding the origin of the emergent hyperbolic geometry of complex networks. On the same time, our  hope is that further research in this direction could also indicate a path for establishing a cross-fertilization between network theory and quantum gravity.
 
\section*{METHODS}
 
 \subsection*{Degree distribution of  growing simplicial complexes with flavor $s$}
In order to derive the degree distribution of the growing simplicial complexes with flavor $s$,  we use the master equation approach \cite{Doro_book,Newman_book}.
It  can be easily shown that  the average number $\tilde{m}_{d}^{[s]}(k)$ of nodes of  degree $k$ that at each time increase their degree by one is given by 
\bea
\tilde{m}_{d}^{[s]}(k)=\frac{d+(d-1+s)(k-d)}{(d+s)t},
\label{pia111}
\eea 
as long as $d+s\neq0$ , i.e. $(d,s)\neq (1,-1)$ for which the growing simplicial complex reduces to a  chain.
To derive the exact degree distribution of the simplicial complex, we consider the  master equation for   the average  number of nodes $N_{d,}^{t,[s]}(k)$ that at time $t$ have  degree $k$ in a growing $d$ dimensional simplicial complex of flavor $s$. The master equation \cite{Doro_book,Newman_book}  for $N_{d}^t(k)$ reads
\bea
N^{t+1, [s]}_{d}(k)-N^{t, [s]}_{d}(k)=\tilde{m}_{d}^{[s]}(k-1)N_{d}^{t,[s]}(k-1)(1-\delta_{k,d})-\tilde{m}_{d}^{[s]}(k)N^{t,[s]}_{d}(k)+\delta_{k,d}
\eea 
with $k\geq d$.\\
Solving this equation we get both exponential and power-law degree distribution. In particular, the degree distribution $P_d^{[s]}(k)$ of growing simplicial complexes of dimension $d$ and flavor $s$ is given for  
for $d+s=1$ by
\bea
P_{d}^{[s]}(k)&=&\left(\frac{d}{d+1}\right)^{k-d}\frac{1}{d+1}, 
\eea
with  $k\geq d$ while for $d+s>1$ it is given by 
\bea
P_{d}^{[s]}(k)&=&\frac{d+s}{2d+s}\frac{\Gamma[1+(2d+s)/(d+s-1)]}{\Gamma[d/(d+s-1)]}\frac{\Gamma[k-d+d/(d+s-1)]}{\Gamma[k-d+1+(2d+s)/(d+s-1)]},
\label{Pksf0b}
\eea
with $k\geq d.$\\
Therefore for $d+s>1$ the degree distribution is scale-free and has a power-law scaling 
\bea
P_{d}^{[s]}(k)\simeq k^{-\gamma_d^{[s]}}
\eea
for $k\gg 1$, with power-law exponent $\gamma_d^{[s]}$
\bea
\gamma_d^{[s]}=2+\frac{1}{d+s-1}\leq 3.
\eea

\subsection*{Community structure of growing simplicial complexes}

The modularity $M$  evaluates the significance of the community structure of a network. 
It is defined \cite{Newman_book} as 
\bea
M=\frac{1}{2L}\sum_{ij}\left(a_{ij}-\frac{k_ik_j}{2L}\right)\delta(q_i, q_j) \ ,
\eea
where  ${\bf a}$ denotes  the adjacency matrix of the network,  $L$ the  total number of links,  and $\{q_i\}$,  where $q_i=1, 2\ldots Q$,  indicates to which community the node $i$ belongs.
Finding the network community structure that optimizes modularity is NP hard. One of the most popular greedy algorithms to find the community structure is the  generalized  Louvain method   \cite{Gen_louvain,Louvain} that is able to determine a  lower bound on the maximum modularity of the network. 
As shown in Table I of the main text the growing simplicial complexes with  $d\geq 2$ are characterized by a large modularity.

\subsection*{Average clustering coefficient   of growing simplicial complexes}
The local clustering coefficient $C_i$ measures the density of triangles passing through node $i$, i.e. 
\bea
C_i= \frac{(a^3)_{ii}}{k_i (k_i-1)}.
\eea
where  ${\bf a}$ denotes  the adjacency matrix of the network, and where $k_i=\sum_j a_{ij}$ is the degree of node $i$.\\
As shown in Table I of the main text the growing simplicial complexes with  $d\geq 2$ are characterized by a large clustering coefficient.

\section*{Acknowledgments}
 \begin{itemize}
 \item 
 G. B. acknowledges interesting discussions with D. Krioukov,  M. A. Serrano, L. Smolin and R. Sorkin.  
 \item 
 This work has been partially supported  by SUPERSTRIPES Onlus. 
G.B. was partially supported by the  Perimeter Institute for Theoretical Physics (PI). The PI is supported by the Government
of Canada through Industry Canada and by the Province of Ontario through the Ministry of Research and Innovation.
\end{itemize}

\section*{Author Contributions Statement}
\noindent G. B.  and C. R.   designed the research,  conducted the research and   wrote the main manuscript text. 
\section*{Additional Information}

 The authors declare that they have no competing financial interests.\\
 
 Correspondence to ginestra.bianconi@gmail.com\\

\newpage

\end{document}